\documentclass[%
 reprint,
superscriptaddress,
 amsmath,amssymb,
 aps, prl
]{revtex4-1}

\usepackage{amsmath}
\usepackage{graphicx}% Include figure files
\usepackage{caption} 
\usepackage{subcaption} 
\usepackage{cleveref}
\usepackage{float}
\usepackage{dcolumn}% Align table columns on decimal point
\usepackage{bm}% bold math
\begin{document}

\preprint{APS/123-QED}

\title{Filming non-radiative population transfer:\\ Time-resolved x-ray scattering near an avoided crossing}

\author{Matthew R. Ware}
\email{mrware@stanford.edu}
\affiliation{Stanford PULSE Institute, SLAC National Accelerator Laboratory, Menlo Park, CA 94025, USA}
\affiliation{Department of Physics, Stanford University, Stanford, California 94305, USA}

\author{James M. Glownia}
\affiliation{LCLS, SLAC National Accelerator Laboratory, Menlo Park, CA 94025, USA}

\author{James P. Cryan}
\affiliation{Stanford PULSE Institute, SLAC National Accelerator Laboratory, Menlo Park, CA 94025, USA}

\author{Robert Hartsock}
\affiliation{Stanford PULSE Institute, SLAC National Accelerator Laboratory, Menlo Park, CA 94025, USA}

\author{Adi Natan}
\affiliation{Stanford PULSE Institute, SLAC National Accelerator Laboratory, Menlo Park, CA 94025, USA}

\author{Philip H. Bucksbaum}
\affiliation{Stanford PULSE Institute, SLAC National Accelerator Laboratory, Menlo Park, CA 94025, USA}
\affiliation{Department of Physics, Stanford University, Stanford, California 94305, USA}
\affiliation{Department of Applied Physics, Stanford University, Stanford, California 94305, USA}

\date{\today}% It is always \today, today,
\begin{abstract}
We show that time-resolved x-ray scattering 
from molecules prepared in a superposition of electronic states moving through an avoided crossing has new features not found in diffraction from the corresponding classical mixed state.
Photoabsorption in molecular iodine at 520 nm produces a superposition of two dipole-allowed nearly degenerate electronic states, which interact due to non-adiabatic coupling.
We show experimental evidence that the mixing of the nuclear wavepackets from the two electronic states at the avoided crossing leads to ultrafast changes in the  angular composition of the scattering pattern.  
This provides a novel means to study transitions in excited molecular systems. 
We reconstruct a movie of the  nuclear probability density arising from this interference.

\end{abstract}

\maketitle

\section{Introduction}

Non-radiative population transfer is a common phenomenon in photoexcited molecular systems.
For example, 
ultraviolet excitation of the nucleotide guanine can relax
via a seam in the potential energy landscape connecting the ground and excited electronic states~\cite{doi:10.1021/ja069176c}.
Another popular example, retinal, 
uses a non-radiative pathway to
isomerize following photoexcitation~\cite{martinez2010physical}.
Population transfer has been identified in x-ray scattering experiments by comparing scattering patterns with simulations of the molecular dynamics.
These simulations model the excitation as a rovibrational wave packet moving on a potential energy surface as it passes through a conical intersection to the ground state~\cite{minitti_imaging_2015}.
Here we find a time-resolved x-ray scattering signature of the interaction of two electronic states excited simultaneously in a molecule, and we use it to image the time and location of non-radiative population transfer between these states at an avoided crossing.

Time-resolved x-ray and electron scattering are powerful techniques for imaging atomic arrangements and motion in molecules.
It has been used to image coherent vibrational oscillations and dissociation in molecular gases, clusters, and solutions \cite{minitti_imaging_2015,glownia_self-referenced_2016,Glownia_reply_2017,kupper_x-ray_2014,Tais_Xenon,van2016atomistic,PhysRevLett.117.013002}. 
Here we show that when the initial wave packet is prepared on two electronic potential surfaces, the scattering can acquire an additional striking feature: an ultrafast change in the symmetry of the scattering pattern as the surfaces interact.  We demonstrate this effect in molecular iodine experiments, where a single laser pulse excites two nearly degenerate electronic states, which subsequently pass through an avoided crossing.

We consider the case of a rovibrational wavepacket excited on two electronic states that are close in energy but have different symmetry. 
The symmetry differences lead to different angular distributions of the the excited states.
We  follow the non-radiative population transfer between these states. 
Both theory and experiments show that the states interact at an avoided crossing, leading to a rapid change in the angular symmetry of the x-ray scattering pattern. 

\begin{figure}
\includegraphics[width=0.7\linewidth]{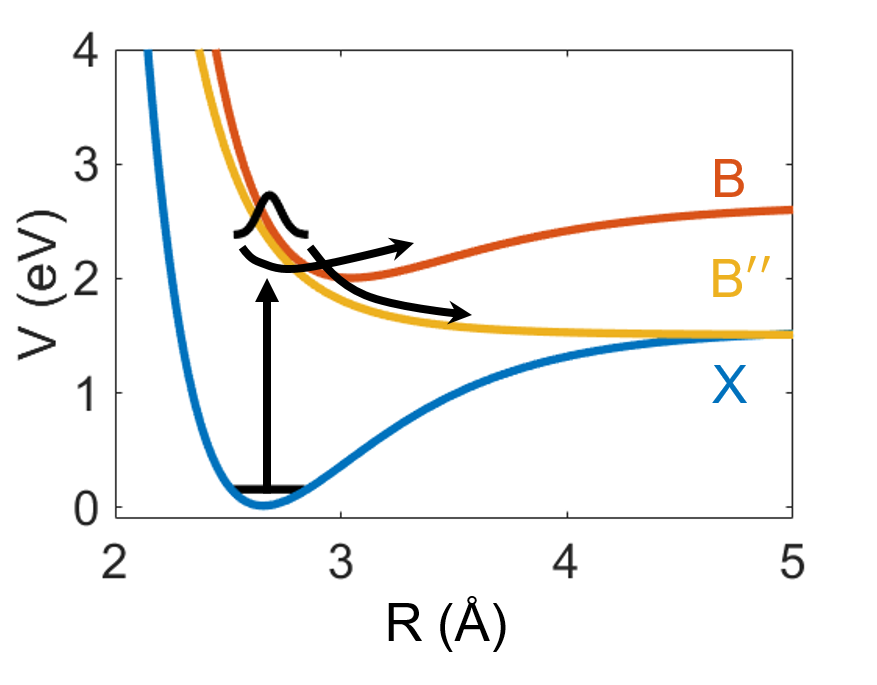}
\captionsetup{justification=raggedright,
singlelinecheck=false
}
\caption{Potential energy diagram of the electronic states in molecular iodine a single 520~nm photon may access.
The X to B transition is a parallel excitation.
The X to B$^{\prime\prime}$ transition is a perpendicular excitation \citep{tellinghuisen_spontaneous_1972,mulliken_iodine_1971}.
The two trajectories highlight the non-adiabatic population transfer near the avoided crossing.} \label{fig:pes}
\end{figure}

Figure \ref{fig:pes} sketches out the non-radiative dynamics in iodine following excitation. Laser-induced resonant photoabsorption in molecular iodine from  the X ($^1\Sigma^+_g$) state at 520nm produces excitation of the bound B ($^3\Pi_{0^+u}$) state and dissociative B$^{\prime\prime}$ ($^1\Pi_u$) state in the ratio 4:1 \cite{tellinghuisen_transition_1982}.  The B and B$^{\prime\prime}$ states can interact non-radiatively through nuclear motion and spin-orbit coupling~\cite{scherer_heterodynedetected_1992,broyer_g_1975,broyer_direct_1975}. 
The initial wavepacket dissociating on the B$^{\prime\prime}$ state can be recaptured in the region of the avoided crossing by making a non-adiabatic transition onto the bound B state.
The non-radiative population transfer between the B and B$^{\prime\prime}$ states leads to interferences between the two nuclear wavepackets.
Because these wavepackets have different initial geometric alignment, this coupling leads to ultrafast changes in the angular symmetry of scattered x-rays or electrons as we now describe.

\section{Theory}  

The time-dependent x-ray scattering intensity may be expressed as 
\begin{equation}
\frac{dI}{d\Omega}= %left side of eqn.
\frac{d\sigma_{Th}}{d\Omega} %Thomson factor
\int d\vec{k} I(\vec{k}) %Integ. I(k)
\frac{\omega}{\omega_s} %inelasticity
% \left< \psi_m(\tau) \left| 
\left< \psi(\tau) \left| % bra
\hat{F}(\vec{Q}) % operator
% \right| \psi_m(\tau) \right>
\right| \psi(\tau) \right> % ket
\label{eqn:gen_scat}
\end{equation}
where 
$\frac{d\sigma_{Th}}{d\Omega}$ 
is the Thomson scattering cross-section, 
$I(\vec{k})$ is the incident x-ray flux,
$\left| \psi(\tau) \right>$ 
is the state of the molecule at pump-probe delay $\tau$, 
$\vec{k}$ is the incident x-ray momentum, $\vec{k}_s$ is the scattered x-ray momentum, $\vec{Q}=\vec{k}-\vec{k}_s$ is the momentum transfer, and 
$\hat{F}(\vec{Q})=
\sum_{j,l}e^{i\vec{Q}\cdot(\vec{r}_j-\vec{r}_l)}$ is the scattering operator,
where $\vec{r}_{j,l}$ 
are the electronic coordinates~\cite{Hau-Riege_2014}. (The formalism for electron scattering is similar, wherein the scattering factor $\hat{F}(\vec{Q})$ must be modified to include the Coulombic interaction between electrons and nuclei, see~\cite{ben1997ultrafast}).

For scattering within the pulse bandwidth (${\omega} \approx {\omega_s}$) and collimated beam input  $I_0$, Equation~\ref{eqn:gen_scat} reduces to the product of three factors:
\begin{equation}
\label{eqn:S(Q)}
\frac{dI}{d\Omega}=
\frac{d\sigma_{Th}}{d\Omega}
I_{0}
S(\vec{Q},\tau)
\end{equation}
where $S(\vec{Q},\tau)$ is a time- and angle-dependent polarization-corrected scattering probability, which arises from the expectation value of the scattering operator $\hat{F}$ in Eqn. \ref{eqn:gen_scat} with respect to the full molecular wavefunction. $S(\vec{Q},\tau)$ may subsequently be expressed as an incoherent sum of the scattering from each electronic state because the coherent cross-terms between electronic states in x-ray scattering are smaller than the incoherent contributions by several orders of magnitude~\cite{bennett_heterodyne-detected_2016}.

The incoherent contribution $S_{N}(\vec{Q},\tau)$  from each electronic state \textit{N} is simplified considerably
by the independent atom approximation so that the scattering probability for a homonuclear diatomic molecule becomes a simple function of the atomic separation $\vec{R}$ \cite{henriksen_theory_2008}:

\begin{equation}
S_N(\vec{Q},\tau) = 
2 \left| f_A(\vec{Q})\right|^2 
\left(1+Re\left\{
\int d\vec{R} 
\rho_N(\vec{R},\tau) 
e^{i\vec{Q}\cdot\vec{R}} 
\right\}\right).
\label{S0}
\end{equation}
Here  
$f_A(Q)$ is the atomic scattering factor and
$\rho_N (\vec{R},\tau)$ 
is the ensemble-averaged nuclear probability density of electronic state \textit{N} at time $\tau$.
The exponential function in Eqn. \ref{S0} can be rewritten using the spherical Bessel expansion:
\begin{equation}
e^{i\vec{Q}\cdot\vec{R}} = \sum_{l} i^l (2l + 1) P_{l}(\cos\theta) j_l(QR).
\label{eqn:bessel_expan}
\end{equation}
where $j_l(QR)$ is the $l$th spherical Bessel function. 
The x-ray diffraction pattern projects onto Legendre polynomials that describe the target probability density \cite{baskin_oriented_2006,lorenz_interpretation_2010}. 
For inversion-symmetric systems, e.g. the nuclear probability density following a dipole excitation, only \textit{even} Legendre polynomials contribute:
\begin{equation}
S_{N}(Q,\theta,\tau)= \sum_{l = 0,2,...} \sqrt{\frac{2\pi}{2l+1}}P_{l}(\cos\theta) S_{N,l}(Q,\tau).
\label{eqn:leg_decomp}
\end{equation}

This analysis suggests that ultrafast transitions between different electronic states can be detected by ultrafast changes in the  Legendre decomposition of the scattering pattern or, rather, the angular symmetry.
This provides a novel means to study transitions in excited molecular systems. 
We now apply these ideas to x-ray scattering in molecular iodine to show how non-radiative transitions may be identified.  

Prior to photoexctiation, the initial iodine molecular ensemble is a thermal equilibrium distribution of rovibrational states on the ground (X) Born-Oppenheimer electronic potential energy surface.  It is both isotropic and time-independent, so prior to photoexcitation, $S_{X}(Q)$ is also independent of time and isotropic.

The signal after photoexcitation is neither 
isotropic nor time-independent. 
Isotropy is broken because excitation probabilities for the X $\rightarrow$ B and X $\rightarrow$ B$^{\prime\prime}$ transitions have an angle dependence of $\cos^{2}\theta$ and $\sin^{2}\theta$, respectively, where $\theta$ is the angle of the interatomic axis with respect to the excitation laser polarization.

The two terms $P_{0}(\cos\theta) = 1$ and $P_{2}(\cos\theta) = (3\cos^2 \theta -1)/2$ are sufficient to describe the scattering following direct single-photon excitation either to B or B$^{\prime\prime}$, whereas the scattering is completely described by the isotropic term $P_0$ prior to excitation.

\begin{figure*}
\captionsetup[figure*]{justification=justified, singlelinecheck=off} 
\centering
  \includegraphics[width=6.0in]{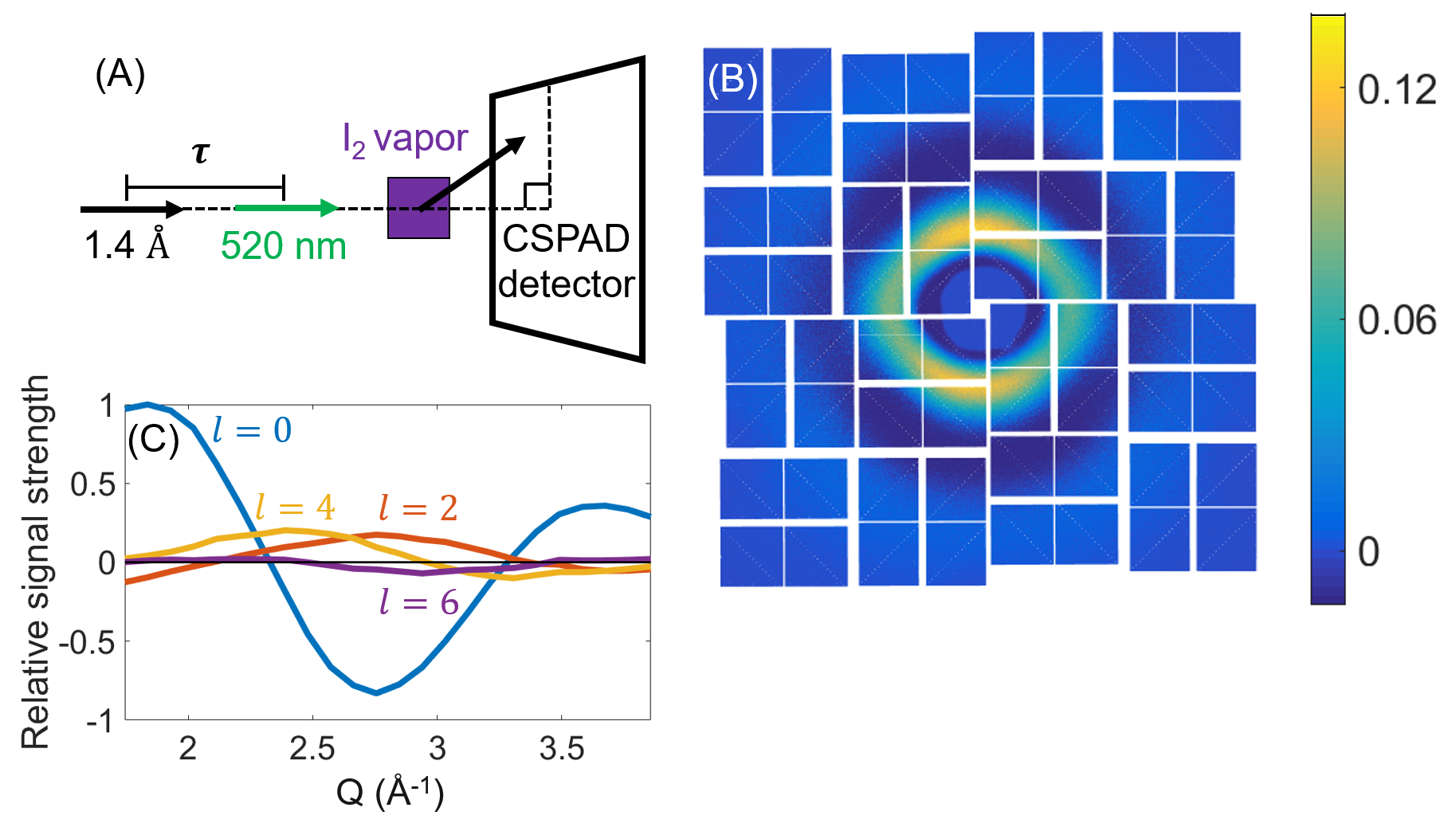}
\caption{
(A) 
Beamline schematic in the XPP hutch at the LCLS. (B) A representative image of the polarization-corrected x-ray scattering signal at the detector for an optical-pump-x-ray-probe delay of 120~fs, with the unpumped x-ray scattering signal subtracted.
The scale is in arbitrary detector units relative to the maximum of $S(\vec{Q},\tau)$ at this delay.
The data are scaled at each radius in Q to account for the normalization over the Ewald’s sphere, $Q^2 \left( S_{pumped}(\vec{Q},\tau)-S_{unpumped}(\vec{Q}) \right)$. (C) We show the projection of the CSPAD image onto the zeroth, second, fourth, and sixth order Legendre polynomials as described in Eqn. \ref{eqn:leg_decomp} in the text.
}
\label{fig:raw}
\end{figure*}

The time dependence of the x-ray scattering pattern following excitation is more complicated, but the independent atom approximation simplifies the analysis by treating each electronic state as an independent potential energy surface in the geometrical space of atomic positions.  The excitation therefore separates into three wave packets with time-dependent ensemble-averaged nuclear probability densities $\rho_{X}(\vec{R},t)$, $\rho_{B}(\vec{R},t)$, and $\rho_{B^{\prime\prime}}(\vec{R},t)$ on the X, B, and B$^{\prime\prime}$ excited state surfaces, respectively:
\begin{equation}
\begin{split}
&S(Q,\theta,\tau)= \\&
S_{X}(Q,\theta,\tau)
+S_{B}(Q,\theta,\tau)
+S_{B{\prime\prime}}(Q,\theta,\tau) =
\\&
P_{0}(\cos\theta)
\left[
S_{X,0}(Q,\tau)
+S_{B,0}(Q,\tau)
+S_{B{\prime\prime},0}(Q,\tau)
\right]+
\\&
P_{2}(\cos\theta)
\left[
S_{X,2}(Q,\tau)
+S_{B,2}(Q,\tau)
+S_{B{\prime\prime},2}(Q,\tau)
\right].
\end{split}
\label{eqn:sum_of_S}
\end{equation}
The $l=0$ and $l=2$ terms in Eqn. \ref{eqn:leg_decomp} thus describe the angular dependence of this scattered x-ray intensity distribution. 
Note that the ground X state Raman excitation is of first order and hence must be included in the above analysis, for example see \cite{lorenz_interpretation_2010}.

Wave packets excited from X to the B and B$^{\prime\prime}$ states pass through an avoided crossing within the first 200~fs according to previous studies~\cite{scherer_heterodynedetected_1992}, when the iodine atoms are separated by about 3~\AA~ as shown in Fig. \ref{fig:pes}.
Non-adiabatic transitions at this point from B to B$^{\prime\prime}$ and B$^{\prime\prime}$ to B are possible.
For example, this could increase the fraction of dissociating molecules, which increases 
$S_{B^{\prime\prime}}(\vec{Q},\tau)$ and decreases  $S_{B}(\vec{Q},\tau)$.
Since these scattering patterns evolve in distinct ways, this change is observable. 
The Legendre decomposition coefficients in Eqn. \ref{eqn:leg_decomp} will change as well, but for a classical mixture of molecules in the B or B$^{\prime\prime}$ state the scattering is still confined to Legendre components with $l = 0$ and 2.

Because each molecule begins in a superposition of the B and B$^{\prime\prime}$ states, the scattering can acquire Legendre projections $l\ge4$ due to the coherent interference of the wavepackets following population transfer at the avoided crossing. 
To show this explicitly, consider that the state of the system may be represented as a 2-component state 
$
\left| a \cos\theta \psi_B + b \sin\theta  \psi_{B^{\prime\prime}} \right>
$. 
For nondegenerate B and B$^{\prime\prime}$ the complex coefficients $a$ and $b$ evolve rapidly in phase, and this has led previous analyses to conclude that the scattering behavior is identical to a classical mixture for most systems~\cite{mukamel_comment}.
However, from the Franck-Condon point to the avoided crossing, the B and B$^{\prime\prime}$ states are nearly degenerate, so the scattering acquires an additional term due to the non-adiabatic coupling between the states:

\begin{equation}
\begin{split}
& \left<\hat{H}^{\prime}\right> =
\left(
  \begin{array}{cc}
  a \cos\theta &  b \sin\theta \\
  \end{array}
\right)
\left(
  \begin{array}{cc}
   0 & H^{\prime} \\
   H^{\prime*} & 0\\
  \end{array}
\right) 
\left(
  \begin{array}{c}
  a \cos\theta \\   
  b \sin\theta \\
  \end{array}
\right) \\&
\propto \cos \theta\sin \theta
\end{split}
\label{eqn:LdotS}
\end{equation} 
In lowest order (the Landau-Zener approximation \cite{LANDAU1977298}) this contributes a term proportional to $\cos^2\theta\sin^2\theta$ to the scattering, which can be decomposed into Legendre terms with $l = 0, 2, 4$. More generally the evolution of a coherent preparation described by $\rho(\vec{R},\tau)$ under this Hamiltonian should lead to additional even orders of $P_l (\cos\theta)$ higher than $l=4$. Hence, the non-radiative transitions in molecular iodine may be isolated by their projection onto Legendre polynomials of order~$l\ge 4$, and more generally, non-radiative transitions which mix wavepackets of different geometric alignment will lead to ultrafast changes in the angular symmetry of the time-resolved scattering.

\begin{figure*}
  \centering
  \includegraphics[width=6.0in]{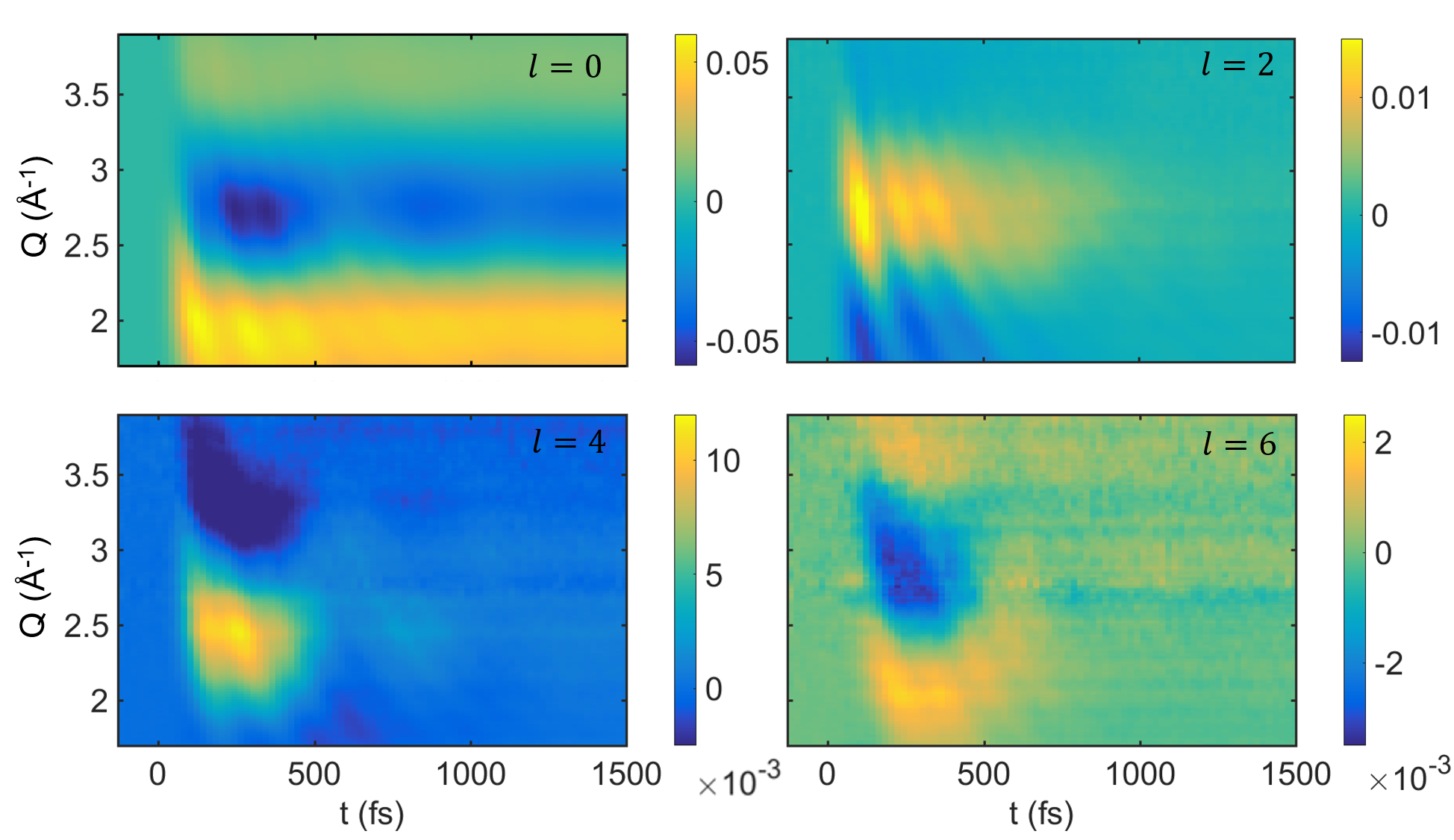}
  \caption{Polarization-corrected difference signal $\Delta S_{l}(Q,\tau)=S_{l,pumped}(Q,\tau)-S_{l,unpumped}(Q)$ projections onto Legendre polynomials, $P_{l}(\cos \theta)$ for $l=0$, $l=2$, $l=4$, and $l=6$.
(See Eqn. \ref{eqn:leg_decomp}).
The $l=0$ and $l=2$ projections are expected from classical considerations, but the $l=4$ and $l=6$ projections are evidence for wave packets in a quantum superposition of B and B$^{\prime\prime}$.  
The signal in $l\geq 2$ decay due to rotational dephasing \cite{glownia_self-referenced_2016}.
Each projection has a relative phase shift due to the spherical Bessel transformation it is associated with.
We have scaled the figures relative to the maximum signal strength in $S_0 (Q,\tau)$.
The sign change from $l=4$ to $l=6$ and the relative peak signal strengths of $l=4$ and $l=6$ are both consistent with the Legendre expansion of $\cos \theta \sin \theta$ from Eqn. \ref{eqn:LdotS}.}
\label{legplots}
\end{figure*}

\section{Experiment}  
To look for evidence of ultrafast symmetry changes following non-radiative population transfer, molecular iodine vapor was resonantly excited from the X state to the B and B$^{\prime\prime}$ states by an ultrafast laser (520$\pm$5 nm, 40 $\mu$J, 50fs, vertically polarized, focused to $\sim$10$^\textrm{12}$ W/cm$^2$) and probed by a spatially coherent beam of 9.0 keV 2 mJ 40 fs horizontally polarized x-rays provided by the LCLS \cite{glownia_self-referenced_2016}.
The copropagating cross-polarized visible and x-ray laser beams were focused into molecular iodine vapor with a
a column density of $\sim$10$^{18}$ cm$^{-2}$.
Approximately 10$^7$ x-rays per pulse were scattered onto a 2.3 megapixel silicon array, the CSPAD \cite{blaj_x-ray_2015}. Figure \ref{fig:raw} shows a representative image of the signal at the detector.
Up to 50 scattered x-rays per pulse per pixel were detected.
The x-rays probe the iodine at a variable time delay following photoexcitation,  producing 
an x-ray scattering signal $dI(Q,\theta,\tau)/d\Omega$ as described by Eqn. \ref{eqn:S(Q)}.

To analyze the data, we rebin the data from $Q_{min} = 0$ to $Q_{max} = 4.0$~\AA using 100 bins, and each bin contains up to 5,000 photons per bin.
Figures~\ref{fig:raw}, \ref{legplots}, and \ref{fig:delay} are scaled relative to this maximum scattering strength. Following the rebinning, we project onto the Legendre polynomials at each radius in momentum transfer, $Q$. The time-dependent signal in these Legendre projections has an associated temporal uncertainty of 40~fs associated with the length of the x-ray pulse and 10~fs due to the jitter uncertainty at the LCLS (following time-tool correction).
The pulse length of 40~fs integrates any periodic features faster than 40~fs to zero, but it does not limit our time resolution for signals with longer period. For transient signals, as shown in Figure~\ref{fig:delay}, the primary source of uncertainty is the machine jitter of 10~fs.

\section{Results and discussion}  

The Legendre decomposition analysis of Eqn. \ref{eqn:leg_decomp} 
is shown in Figure \ref{legplots}. The unpumped x-ray scattering signal has been subtracted to remove detector imperfections and background.  In addition to the strong expected contributions to scattering from $l=0$ and $l=2$ for a classical mixture of molecules in the B or B$^{\prime\prime}$ states, we see clear contributions from $l=4$ and, additionally, a smaller but distinct contribution from $l=6$. We now argue that this arises from the non-radiative mixing of wavepackets with different initial alignments.

We have verified this signal is physical and not a detector artifact or issue with projecting the data. The $l > 2$ signal disappears if we change the wavelength to 800nm, where the system does not encounter an avoided crossing. Also the signal strength of the higher order Legendres is reduced by $1/4$ if we use elliptically polarized light instead of linearly polarized light, indicating this phenomenon relies on the linear alignment of the two excited state populations. Moreover, each Legendre order of the data in Figure \ref{legplots} has an associated phase shift consistent with the phase-shift between the associated spherical Bessel functions from Eqn. \ref{eqn:bessel_expan} at low Q.

The selection rules for single-photon excitation cannot account for $l\geq 4$, suggesting some additional interaction.  There are two candidates:
(1) Higher order (multiphoton or Raman) terms left out of the photoexcitation description lead to higher powers of $\cos\theta$ and $\sin\theta$ in $\rho(\vec{R},t)$. 
Or (2) 
the non-radiative coupling shown in Eqn. \ref{eqn:LdotS} 
leads to cross-terms that project onto all even orders of Legendre polynomials. 

(1) In the case of multiphoton excitations, we would anticipate the ensemble to have an initial distribution like $\cos^{2n} \theta \sin^{2m} \theta$ at $t_0$, which would project at most onto $P_{2n+2m}(\cos \theta)$ at $t_0$ and the subsequent dephasing would decrease the order of this projection. Here n is the number of parallel transitions along the laser polarization, and m is the number of perpendicular transitions.

(2) In the case of the non-radiative coupling described in Eqn. \ref{eqn:LdotS}, the x-ray diffraction would only project onto $l=0$ and $2$ at the time of excitation. After a short delay, the system can project onto $l \geq 4$ following mixing of the wavepackets at the avoided crossing.

\begin{figure}
\includegraphics[width=0.9\linewidth]{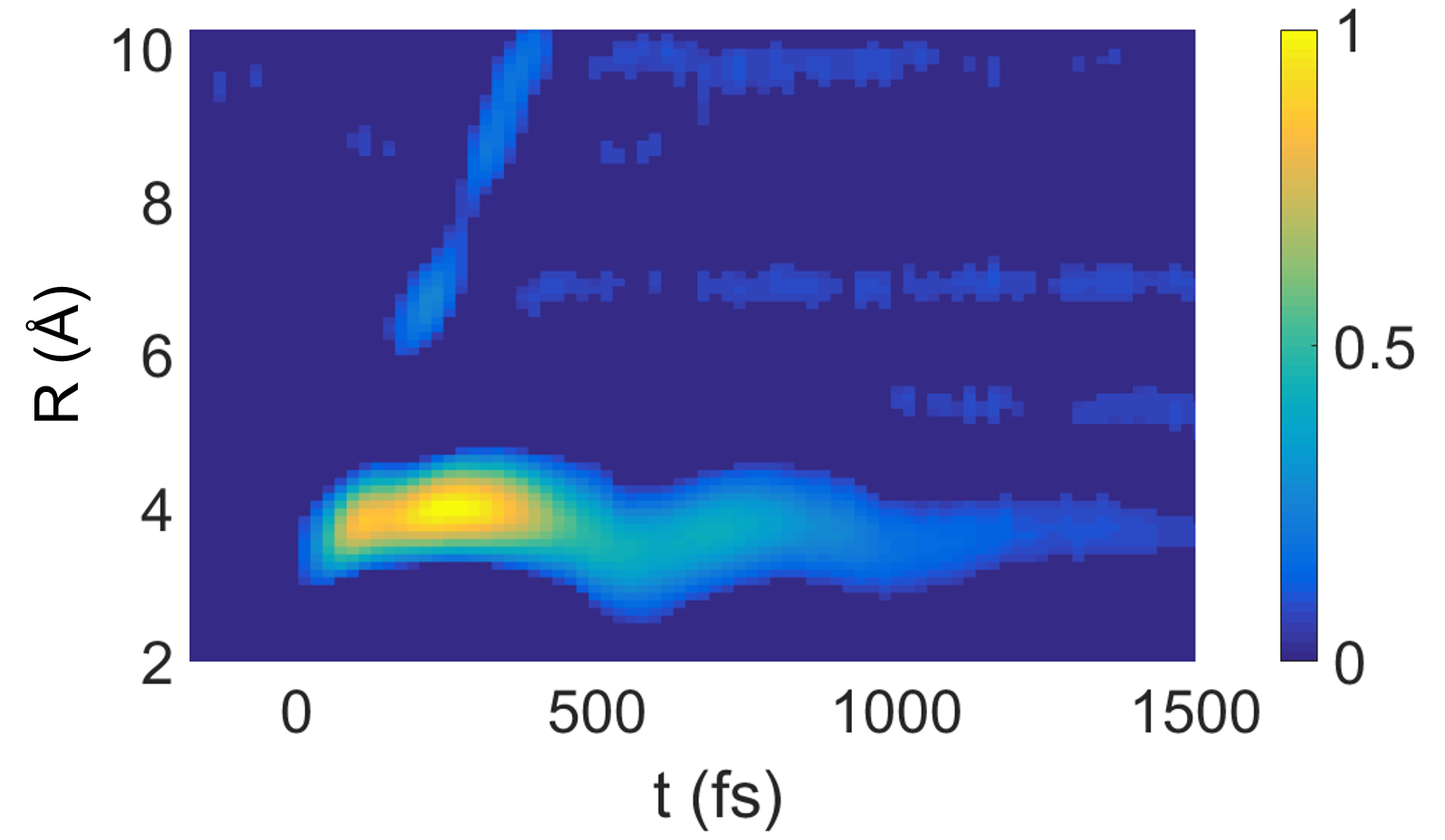}
\captionsetup{justification=raggedright,
singlelinecheck=false
}
\caption{Real-space reconstruction of the Legendre 4 projection of the x-ray diffraction data.
Legendre 6 produces the same features with reduced fidelity.
We observe a dissociation along the B$^{\prime\prime}$ state of velocity $18.6\pm 2.2$ \AA/ps and periodic bound state motion (T=550 fs) within the B state.
The vibrational motion clearly shows the delayed onset, appearing at 3\AA, 50~fs after the excitation.
The dissociation signal from $4.5-6$ \r{A} at 100 fs disappears due to the limited Q range of the experiment, and the horizontal lines are artifacts of the reconstruction due to the limited Q range. 
} 
\label{fig:recon}
\end{figure}

The higher order Legendre components of the scattering data have two distinctive features that help us to understand their origin: 

$\bullet$ The real-space reconstruction of the $l=4$ and $l=6$ Legendre projections unveil dynamics consistent with the B and B$^{\prime\prime}$ states.
We show the $l=4$ reconstruction in Fig. \ref{fig:recon}. 
To recover the real-space dynamics, we invert the associated spherical Bessel transformations from Eqn. \ref{eqn:leg_decomp}.
(For diatomics, the inverse spherical Bessel transformation reconstructs the nuclear charge density. For molecules with more than two atoms, the transformation produces the pair-correlation functions  \cite{lorenz_interpretation_2010}.)
We identify the bound B state by its period of $T=550$~fs about an equilibrium position near 3~\AA, and we identify the dissociation on the B$^{\prime\prime}$ state by the dissociation velocity of $18.6\pm 2.2$~\AA/ps, consistent with the dissociation velocity estimate of 16.7~\AA/ps from the kinetic energy release. 
% We therefore see no indication of (1) Raman excitation on the ground state nor higher lying electronic states above the B states. We do see a clear indication that (2) the B and B$^{\prime\prime}$ states are responsible for the time-delayed $l\ge 4$ signal.
We conclude that the $l=4$ signal is consistent with (2), ie. the dynamcs on the B and B$^{\prime\prime}$ following non-radiative population transfer.
% the B and B$^{\prime\prime}$ states are responsible for the time-delayed $l\ge 4$ signal.

\begin{figure}
\includegraphics[width=0.9\linewidth]{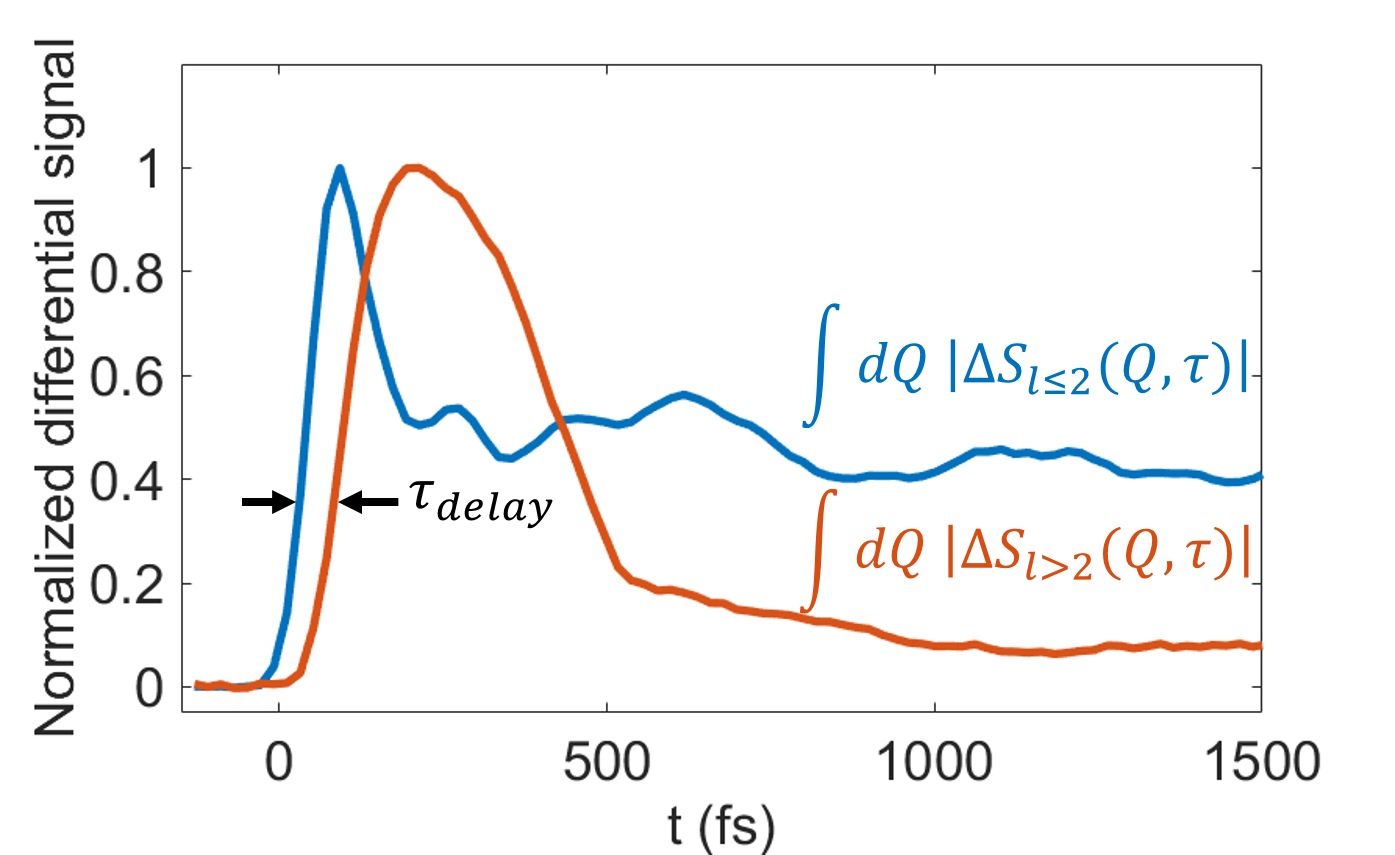}
\captionsetup{justification=raggedright,
singlelinecheck=false
}
\caption{The higher order Legendre projections are delayed by $\tau_{delay}=50\pm 20$~fs.
This signal is consistent with evolution via non-adiabatic and spin-orbit coupling at the avoided crossing.} \label{fig:delay}
\end{figure}

$\bullet$ The integrated signal of each $l\ge4$ projection is delayed by 50$\pm$20~fs with respect to the $l\le2$ Legendre projections as shown in Figure \ref{fig:delay}.  
This delayed response is consistent with (2) the upper limit of the crossing, $t_{crossing}\le 200$~fs, as given by earlier experiments~\cite{scherer_heterodynedetected_1992}. The different initial starting position is also reflected in the oscillating wave packet in Fig. \ref{fig:recon} which begins at 3~\AA and not near the Franck-Condon region around 2.7~\AA.

Therefore, this ultrafast change in the angular symmetry of the x-ray scattering from iodine from an initial Legendre composition of $l=0$ and 2 to a Legendre composition including $l\ge4$ is most likely due to the known non-radiative coupling between the B and B$^{\prime\prime}$ states. Moreover, for molecular systems that couple wavepackets of different initial alignment through non-radiative coupling, this result implies that an associated ultrafast change in the angular symmetry of the scattering may be observed.

\section{Conclusion}  
% We conclude that the $l\ge4$ scattering distributions result from the interference of the B and B$^{\prime\prime}$ wavepackets following mixing at the avoided crossing. After population transfer onto the B state through spin-orbit coupling, the initially dissociative state becomes bound.
% Likewise, some of the initial B state population transfers to the dissociative B$^{\prime\prime}$ state. 

% In summary, we have observed the  transfer of population between coherently prepared B and B$^{\prime\prime}$ wave packets in iodine as they pass through an avoided crossing and have reconstructed the subsequent dynamics arising from their quantum interference.
% We see the coherent interference on both the bound B state and the dissociative B$^{\prime\prime}$ state through the time-delayed, higher order Legendre projections of the x-ray diffraction.
% This also demonstrates that the same mechanism which leads to predissociation of a bound state also leads to preassociation, i.e. preemptive capture of a dissociating molecule.

The previously studied predissocation of iodine has provided us with an opportunity to demonstrate that non-radiative population transfer coincides with an ultrafast change in the angular symmetry of the time-resolved scattering under certain conditions.
The key condition is that the non-radiative population transfer must mix states of different initial alignment to the external laser field.
In the case of molecular iodine, two degenerate electronic states were excited: one with a molecular axis initially parallel to the external field and another with a molecular axis initially perpendicular to the external field.
However, there is no reason for the coupled states to both be photoexcited.
We would anticipate a similar phenomenon in a molecular system that couples the geometrically aligned excited state to the isotropic ground state.
In such a system, the initial alignment, say, $|\cos \theta|^2$ would interfere with the ground state to produce an angular distribution of $\cos \theta$ which projects onto all orders of Legendre polynomials.

We conclude that, for molecular systems which non-radiatively mix populations of different symmetries, that an ultrafast change in the angular symmetry of the scattering is a key observable for future experimental work.

We wish to acknowledge the Glownia L2816 collaboration for access to data from their LCLS experiment, which were used in this analysis, and we wish to acknowledge useful discussions with Tais Gorkhover in the preparation of this letter. 
This research is supported through the Stanford PULSE Institute, SLAC National Accelerator Laboratory by the U.S. Department of Energy, Office of Basic Energy Sciences, Atomic, Molecular, and Optical Science Program.
Use of the Linac Coherent Light Source (LCLS), SLAC National Accelerator Laboratory, is supported by the U.S. Department of Energy, Office of Basic Energy Sciences under Contract No. DE-AC02-76SF00515. Matthew Ware is supported by the Stanford Graduate Fellowship.

\bibliography{Filming_predissociation_of_iodine_v4}

\end{document}